\documentclass[prd,floatfix,showpacs,11pt]{revtex4}

\usepackage{amssymb}
\usepackage{amsmath}
\usepackage[dvips]{graphicx}
\usepackage[dvips]{color}

\DeclareMathOperator{\tr}{tr}

\newcommand{\be}{\begin{equation}}
\newcommand{\ee}{\end{equation}}
\newcommand{\bse}{\begin{subequations}}
\newcommand{\ese}{\end{subequations}}
\newcommand{\eq}[1]{Eq.~(\ref{#1})}
\newcommand{\eqs}[1]{Eqs.~(\ref{#1})}

\newcommand{\RR}{\mathbb{R}}

\newcommand{\ZZ}{\mathbb{Z}}

\newcommand{\SU}{{\rm SU}}

\newcommand{\su}{{\mathfrak{su}}}
\newcommand{\so}{{\mathfrak{so}}}
\newcommand{\inv}{^{-1}}
\newcommand{\dparc}[2]{\displaystyle\frac{\partial #1}{\partial#2}}
\newcommand{\n}{p}

\begin{document}

\title{New self-dual solutions of $\SU(2)$ Yang-Mills theory in Euclidean Schwarzschild space}

\author{Ricardo A. Mosna}
\email{mosna@ime.unicamp.br}
\affiliation{Instituto de Matem\'atica, Estat\'\i stica e Computa\c{c}\~ao Cient\'\i fica,
Universidade Estadual de Campinas, 13083-859, Campinas, SP, Brazil.}

\author{Gustavo Marques Tavares}
\email{gmt@ifi.unicamp.br}
\affiliation{Instituto de F\'\i sica Gleb Wataghin,
Universidade Estadual de Campinas, 13083-970, Campinas SP, Brazil.}

\date{\today}

\begin{abstract}
We present a systematic study of spherically symmetric self-dual solutions of $\SU(2)$ Yang-Mills theory on Euclidean Schwarzschild space.
All the previously known solutions are recovered and a new one-parameter family of instantons is obtained.
The newly found solutions have continuous actions and interpolate between the classic Charap and Duff instantons.
We examine the physical properties of this family and show that it consists of dyons of unit (magnetic and electric) charge.
\end{abstract}

\maketitle


\section{Introduction}

The study of $\SU(2)$ instantons, i.e., finite action smooth (anti) self-dual solutions of the $\SU(2)$ Yang-Mills (YM) equation, has been extensively carried out in Euclidean flat space, and numerous interesting properties of them are by now well established~\cite{shifman}.
These solutions correspond to global minima of the action in each topological sector and, therefore, are associated with the leading terms in semiclassical approximations to the path integral. In particular, instantons are associated with tunneling between degenerate vacua in Yang-Mills theory at zero temperature.

At the classical level, it is well known that these solutions are classified by $\pi_3(S^3)=\ZZ$ and that their action, which coincides with the first Pontryagin class, is restricted to integer values~\cite{nakahara}. These results are related to the topology of flat space and are not expected to hold in manifolds with other topologies.

One important feature of self-dual solutions in Euclidean backgrounds (not only in flat space) is that the associated energy-momentum tensor vanishes and thus such field configurations do not disturb the geometry of spacetime~\cite{cd}. Hence, given a solution of Einstein's equation in vacuum and a self-dual solution of YM theory in such background, we immediately have a full solution of Einstein-Yang-Mills theory. Therefore, solutions of Einstein's equation in vacuum provide the most interesting manifolds (from a physical viewpoint) to study self-dual solutions of YM theory.

Furthermore, although gravity is expected to have a weak effect on the perturbative sector of field theories, it might drastically change the nonperturbative sector. This is not a surprise since gravity can change the topology of spacetime and therefore strongly affect the configuration space for the fields. For example, a number of works have considered self-dual solutions of Yang-Mills theory in various Euclidean backgrounds and found solutions with no counterpart in flat space \cite{cd,cd1,bjcc,radu,radu2,ads,kim}. Even the usual tunneling interpretation does not hold in some spacetime geometries~\cite{tekin}. Another remarkable difference is that the action can take continuous values as has been exhibited for the AdS, Schwarzschild and other spherically symmetric spacetimes~\cite{radu,ads}.

In this work we approach the question of determining, in a systematic way, the possible spherically symmetric
smooth self-dual solutions of Yang-Mills theory in the Euclidean Schwarzschild background.
As a result, we recover all the previously known solutions and, moreover, disclose a whole new one-parameter family
of instantons with continuous actions. This family interpolates between the classic Charap and Duff's solutions~\cite{cd,cd1}
with actions $1$ and $2$. We examine the physical properties of this family and show that
it consists of dyons of unit (magnetic and electric) charge.
Like the previously obtained instantons in this background, these new solutions are associated with
a constant gauge-invariant static potential (no tunneling).

The structure of this paper is as follows. We first discuss the spherically symmetric ansatz employed by us, paying special attention
to regularity issues that naturally arise in this context. We then prove the existence of the aforementioned solutions, study their global
behavior and examine their physical properties.
Finally, we employ the recently introduced mapping method~\cite{mapping,mapping2} to obtain a new variable
according to which these solutions can be analytically expressed as a power series with apparently infinite radius of convergence.
We close with some general remarks which include a brief discussion on the possible mathematical relevance of our results to the study
of instantons in asymptotically locally flat (ALF) geometries~\cite{insgrav}.

\section{Ansatz}

The Euclidean Schwarzschild space is a noncompact complete Riemannian manifold topologically given by $M = S^2 \times \RR^2$,
and endowed with the following metric, defined on the open set $S^2 \times (\RR^2\backslash \{ 0 \})\subset M$:
\be
\label{metric}
ds^2 = \left( 1-\frac{2m}{r} \right) d\tau^2 + \left( 1-\frac{2m}{r} \right)\inv dr^2 + r^2 d\Omega ^2.
\ee
In the above expression $m>0$, $d\Omega^2$ is the usual round metric in $S^2$, and
$(r, \tau)$ are polar coordinates on $\RR^2\backslash \{0\}$,
which take values on the intervals $(2m,\infty)$ and $[0,8 \pi m)$, respectively.
It is well known that the metric~(\ref{metric}) can be uniquely extended to the
whole $M$, and that such extension is a nonsingular solution
of the vacuum Einstein equations, with nontrivial Euler characteristic $\chi=2$.

Using coordinates $x^\mu = (\tau, r \sin \theta \cos \phi, r \sin \theta \sin \phi, r\cos \theta)$, with $(\theta,\varphi)$ the usual angular coordinates in $S^2$, we will adopt the general spherically symmetric ansatz for the gauge potential~\cite{tekin,witten}:
\be
\label{witten ansatz}
A = \frac{\sigma_a}{2i} \left[ \phi(\tau,r) \frac{x^a}{r} d\tau + \beta(\tau,r) \frac{x^a x^c}{r^2} dx^c +  (\gamma(\tau,r) -1)\epsilon_{abc} \frac{x^b}{r} d\left( \frac{x^c}{r} \right) + \alpha(\tau,r) d\left(  \frac{x^a}{r} \right) \right],
\ee
where $\sigma_a$ are the Pauli matrices and $\epsilon_{abc}$ is
the totally antisymmetric tensor, with $\epsilon_{123}=+1$.

Because of the angular character of the time coordinate, care must be taken with the boundary conditions to ensure regularity of the gauge potential. Clearly all of the ansatz functions must be periodic in time.
Also, as $r\rightarrow 2m$, $\phi$ must vanish and all the other functions must become time-independent.

Under a gauge transformation $\tilde A = U^{-1} A U + U^{-1}dU$, with $U=\exp(-i f(\tau,r)x^a \sigma_a/2r)$, the form of the ansatz is unchanged
(this corresponds to a $U(1)$-symmetry of the problem~\cite{tekin,witten}) and this can be used to choose a gauge in which $\beta = 0$.
If we are interested in time-independent self-dual solutions~\cite{fn1} we can also set $\gamma=0$.
The (singular) gauge transformation given by
$U = \exp(-i\varphi\sigma_3/2) \exp(-i\theta\sigma_2 /2)$
then leads to the ansatz~\cite{radu}
\be
\label{ansatz}
A=\alpha(r) \, d\theta \, \frac{\sigma_1}{2i} +
\alpha(r)\sin\theta \, d\varphi \, \frac{\sigma_2}{2i} +
\left( cos\theta \, d\varphi + \phi(r) \, d\tau \right) \, \frac{\sigma_3}{2i}.
\ee
Note that given a singular potential in the form of \eq{ansatz} one can turn it back to the form of \eq{witten ansatz} and get a nonsingular potential as long as we ensure that the regularity constraints are obeyed.

For a potential given by the above expression the self-duality equations are:
\bse
\label{EDPs_acopladas}
\begin{align}
\alpha'(r) &= -\left(1-\frac{2m}{r}\right)^{-1}\alpha(r) \, \phi(r), \label{EDPs_acopladas.a}\\
\phi'(r)   &= \frac{1-\alpha^2(r)}{r^2}. \label{EDPs_acopladas.b}
\end{align}
\ese
It immediately follows that $\phi$ is governed by the ordinary differential equation
\be
\label{EDO_phi}
\frac{r}{2}(r-2m)\frac{d^2\phi}{dr^2}+(r-2m)\frac{d\phi}{dr}-\phi
+r^2\frac{d\phi}{dr}\phi =0,
\ee
and that $\alpha$ is given, in terms of $\phi$, by
\be
\label{alpha}
\alpha^2(r)=1-r^2\phi'(r).
\ee
Notice that this imposes a restriction on the solutions of~\eq{EDO_phi}, since $\alpha$ needs to be a real function
($\alpha^2(r)\ge0$) in order that $A$ be $\su(2)$-valued~\cite{fn2}.

The Lagrangian associated with these solutions is easily calculated,
\be
\label{lagrangian}
\frac{1}{8 \pi^2} \tr (F\wedge \star F) = - \frac{1}{8 \pi^2}\dparc{}{r}\left[ (1-\alpha^2) \phi \right] \; d\tau\wedge dr\wedge d\Omega,
\ee
and using this expression one finds that the action is given by
\be
\label{energia1}
S = - \frac{1}{8 \pi^2} \int_M \tr (F\wedge \star F) = 4 m \left[ (1-\alpha^2) \phi  \right]\Big{|}^{\infty}_{2m}.
\ee

The solutions we will be interested in satisfy $\alpha(2m)=0$ and $\alpha(r)\rightarrow 0$ as $r$ goes to infinity (see section~\ref{section:solutions}).
In this case the action depends only on the asymptotic values of $\phi(r)$:
\be
\label{energia}
S =4m\left[\phi(\infty)-\phi(2m)\right].
\ee
We note that this expression is not gauge invariant:
its simple form is a direct consequence of the gauge choice associated with ansatz~(\ref{ansatz}).

Following the approach described in~\cite{ad} (see also~\cite{thooft,tekin}) one can define an Abelian field strength given by
\[ f_{\mu\nu} = -i \tr \left[\sigma_3 \, (\partial_\mu A_\nu - \partial_\nu A_\mu)\right]. \]
This can be identified with an electromagnetic field and, in particular we find that
\bse
\label{EM field}
\begin{align}
    f^{0i}       & = \phi'(r)\frac{x^i}{r}, \label{electric field} \\
    \star f^{0i} & = \frac{x^i}{r^3}. \label{magnetic field}
\end{align}
\ese
Therefore it is clear that all configurations given by ansatz~(\ref{ansatz}) have unit magnetic charge. The electric charge depends on the asymptotic behavior of $\phi(r)$.

\section{Solutions}
\label{section:solutions}

\subsection{General discussion}

Three solutions immediately arise from~\eqs{EDPs_acopladas}, which are the classic Charap and Duff solutions~\cite{cd}:
\begin{enumerate}
\item[i.] $\phi(r)=0$, $\alpha(r)=1$. In this case, $S=0$.

\item[ii.] $\phi(r)=-\frac{m}{r^2}$, $\alpha^2(r)=1-\frac{2m}{r}$. In this case, $S=1$.

\item[iii.] $\phi(r)=c-\frac{1}{r}$, where $c$ is a constant, and $\alpha(r)=0$. In this case, $S=2$.
\end{enumerate}
We see from~\eqs{EM field} that the solutions with actions $1$ and $2$ above correspond to a monopole and a dyon, respectively~\cite{tekin}. The third solution is moreover Abelian, since
it lies in a fixed $\mathfrak{u}(1)$ inside $\su(2)$ as can be seen from~\eq{ansatz} (see also~\cite{gh}).

It is evident that the second and third (for general $c$) solutions above do not satisfy the regularity condition $\phi(2m)=0$. Going back to Eq.~(\ref{witten ansatz}) we can see that given a gauge potential with $\phi(2m) = \lambda \neq 0$ we can turn it into a regular potential by performing a singular gauge transformation with $U = \exp (i\lambda t \, x^a \sigma_a /2r)$, as long as $8\pi m \lambda = 2\pi p$, with $p \in \ZZ$, and $\alpha(2m) = 0$. This is clearly the case for the aforementioned solutions if we choose $c$ properly. Notice that, unless $\alpha(r)=0$, the nonsingular solution will be time-dependent.

A straightforward application of the power series method and the aforementioned condition on $\phi(2m) = \tfrac{p}{4m}$ restrict
the possible solutions of~(\ref{EDO_phi}), in a neighborhood of $2m$, to $\phi(r)=\sum_{k=0}^\infty a_k (r-2m)^k$,
with $a_0 = \tfrac{p}{4m}, p\in\ZZ$.
It turns out that only the Abelian solution is possible for $p>0$.
For $p \leq 0$ a much more interesting situation arises.
In this case, the coefficient $a_{-p+1}$ is not fixed by the condition $a_0 = \frac{\n}{4m}$,
and the remaining coefficients $a_k$ are determined by a
complicated nonlinear recurrence relation that depends on all the coefficients with index less than $k$
(and hence on $a_{-p+1}$). As a result, for each $p \leq 0$, the free parameter $a_{-p+1}$ gives rise
to a whole family of solutions of~(\ref{EDO_phi}).

We have strong numerical evidence that, for $p\leq -2$, the only possible solution with finite action is again the Abelian solution:
all other possible solutions of \eq{EDO_phi} either diverge as $r\rightarrow \infty$ or render $\alpha$ imaginary.
In this way, we turn our attention to the two remaining possibilities, namely $\n=-1$ and $\n=0$.
These two cases have solutions with very dissimilar behaviors.
In fact, as $r\to 2m$, $\alpha(r)$ goes like $(r-2m)^{1/2}$ and $const + (r-2m)$ for $\n=-1$ and $\n=0$,
respectively. More to the point, the solutions associated with $\n=-1$ and $\n=0$ are not gauge equivalent
(except for the trivial Abelian case). This can be easily seen by plugging the power series
expressions for these solutions into the gauge-invariant expression~(\ref{lagrangian}).
It is not difficult to show that the case $\n=0$ exactly corresponds to the results obtained in~\cite{radu},
in which the authors present a family of solutions with action ranging from $0$ to $2$.
Therefore, in this work we focus on the remaining possible choice ($\n=-1$) which, as discussed above,
leads to time-dependent solutions in the appropriate nonsingular gauge.

It follows from the series solution that $a_1 = \phi'(2m)=\frac{1}{4m^2}$. Writing $s=r-2m$, the first few terms
of the corresponding solution are given by
\be
\label{phi_kappa}
\phi_\kappa(s+2m)=-\frac{1}{4m}+\frac{s}{4m^2} +\frac{\kappa}{16m^3} \, s^2
-\frac{\kappa+1}{16m^4} \, s^3 - \frac{\kappa^2 - 7\kappa -10}{256m^5} \, s^4 + \mathcal{O}(s^5),
\ee
where we introduced the free dimensionless parameter $\kappa$ given by $\kappa=16m^3\, a_2$.
It is not hard to show, directly from the recurrence relation satisfied by $a_n$, that:
\begin{enumerate}
\item  for $\kappa=-3$, $\phi_\kappa$ reduces to $\phi_{-3}(r)=-\frac{m}{r^2}$, which is
the aforementioned Charap and Duff's solution with  $S=1$;

\item  for $\kappa=-2$, $\phi_\kappa$ reduces to $\phi_{-2}(r)=\frac{1}{4m}-\frac{1}{r}$, which is
the aforementioned Charap and Duff's solution with $S=2$.
\end{enumerate}
Unfortunately, the radius of convergence of~(\ref{phi_kappa}) is, in general,
not greater than $2m$.
In this way, this method fails to provide any insight about $\phi(r)$ for $r\gg1$.
In particular, it yields no information whatsoever on the finiteness
(not to say the value) of the action associated with each $\phi_\kappa$ (\eq{energia}).
This kind of question requires considerations on the global behavior of these solutions
and will be elucidated in the last part of this section by studying some analytic properties shared by them.
Then, in section IV, we will use a mapping method
to obtain a series solution that has an infinite radius of convergence.

\begin{figure}[htbp]
\begin{center}
\includegraphics[width=0.6\linewidth]{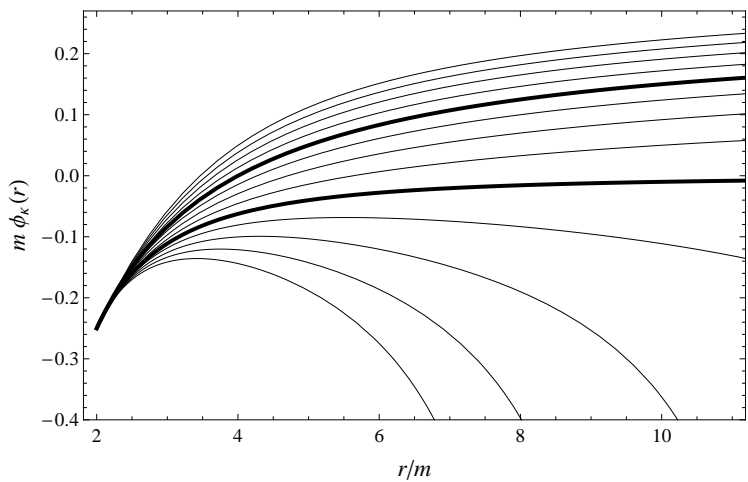}
\caption{Plots of $\phi_\kappa(r)$ for evenly spaced values of $\kappa$ going from $-4$ (lowest curve)
to $-1$ (uppermost curve). The thick curves correspond to the Charap and Duff solutions
($\kappa=-3$ for the lower thick curve and $\kappa=-2$ for the upper thick curve).}
\label{fig:superfamilia}
\end{center}
\end{figure}

Despite its aforementioned limitations, \eq{phi_kappa} can be used to generate initial values
$\left(\phi_\kappa(2m+\epsilon),\phi'_\kappa(2m+\epsilon)\right)$
for~(\ref{EDO_phi}), with $\epsilon$ positive and arbitrarily small.
Such initial values can then be employed to numerically integrate~(\ref{EDO_phi})~\cite{fn3}.
Figure~\ref{fig:superfamilia} shows representative solutions obtained
in this way. The thick curves correspond to the classic Charap and Duff solutions~\cite{cd}.

Figure~\ref{fig:superfamilia} suggests that:
\begin{enumerate}
\item solutions with $\kappa<-3$ (below the lower thick curve) are not limited, and therefore
      do not have finite action (cf \eq{energia});

\item solutions with $-3<\kappa<-2$ (between the thick curves) interpolate between the
      Charap e Duff classic solutions;

\item solutions with different values of $\kappa$ do not intersect.
\end{enumerate}
In what follows, we show that the solutions of interest are indeed those with $-3\le\kappa\le\-2$.
In this case, the corresponding $\phi_\kappa$ does give rise to pairs
$(\phi,\alpha)$ (cf \eq{ansatz}) associated with solutions that interpolate
between Charap and Duff's instantons with actions $1$ and $2$. We also show that item~3 above is true
as long as $\kappa\le-2$, which is precisely the condition for $\alpha(r)$,
as defined by \eq{alpha}, to be real-valued.

\subsection{Global behavior of the solutions}

We start by noting a useful local property of the solutions $\phi_\kappa$. Let $\kappa_1,\kappa_2\in\RR$. If $\kappa_1<\kappa_2$, then, for $r-2m \ll 1$:
\bse
\label{ssmall}
\begin{align}
\phi_{\kappa_1}(r)  < \phi_{\kappa_2}(r),   \label{ssmall.a}\\
\phi'_{\kappa_1}(r) < \phi'_{\kappa_2}(r),   \label{ssmall.b}
\end{align}
\ese
which follows directly from Eq.~(\ref{phi_kappa}).
Therefore plots of $\phi_\kappa(r)$ (as in Fig.~\ref{fig:superfamilia}) corresponding to different
values of $\kappa$ do not intersect for $r$ sufficiently close to $2m$.
A careful analysis of the ordinary differential equation (ODE)~(\ref{EDO_phi}) leads to a much stronger statement.
In fact, we show in the Appendix that for $\kappa_1<\kappa_2\le-2$:
\bse
\label{slarge}
\begin{align}
\phi_{\kappa_1}(r)  < \phi_{\kappa_2}(r)  \quad \forall r\in (2m,\infty),  \label{slarge.a}\\
\phi'_{\kappa_1}(r) < \phi'_{\kappa_2}(r)  \quad \forall r\in (2m,\infty).   \label{slarge.b}
\end{align}
\ese
It follows from~\eq{slarge.a} that plots of $\phi_\kappa(r)$ corresponding to different
values of $\kappa$ do not intersect for any $r\in (2m,\infty)$ as long as $\kappa\le-2$~\cite{fn4}.
We also note that taking $\kappa_2=-2$ in~\eq{slarge.b} leads to $\phi'_\kappa(r) \leq 1/r^2$
whenever $\kappa<-2$, which is precisely the condition needed to guarantee that $\alpha$ is real.

Consider now a fixed value of $\kappa\in(-3,-2)$.
A direct consequence of the above discussion is that the associated action $S_\kappa$ is finite in this case.
Indeed, using \eq{slarge.b} with $\kappa_1 = -3$, we see that $\tfrac{2m}{r^3}<\phi'_\kappa(r)$, and hence
$\phi_\kappa$ is a strictly increasing function. Moreover, it follows directly from \eq{slarge.a} that
\[-\tfrac{m}{r^2}<\phi_\kappa(r) < \tfrac{1}{4m} - \tfrac{1}{r} \, , \qquad \text{for} \, -3<\kappa<-2 ,\]
and thus $\phi_\kappa(r)$ is limited. Therefore, the limit
\be
\label{Ckappa}
C_\kappa:=\displaystyle\lim_{r\to\infty} \phi_\kappa(r)
\ee
exists and $C_\kappa\in (0,\tfrac{1}{4m}]$ for $-3<\kappa\leq  -2$. On the other hand,
it is easy to show, directly from \eq{EDO_phi}, that the asymptotic behavior of $\phi_\kappa(r)$,
also for $-3<\kappa\leq  -2$, is given by
\be
\label{asymptotic}
\phi_\kappa(r) \rightarrow C_\kappa -\frac{1}{r} +\lambda\frac{e^{-2C_\kappa r}}{2C_\kappa}
\qquad (r\gg 1)
\ee
for some $\lambda \in \RR$, which implies $\alpha \rightarrow 0$.
This shows that the action can be calculated from Eq.~(\ref{energia}) and that it is indeed finite. It is also clear from \eqs{electric field} and (\ref{asymptotic}) that these solutions have unit electric charge and therefore correspond to dyons.

\begin{figure}[htbp]
\begin{center}
\includegraphics[width=0.6\linewidth]{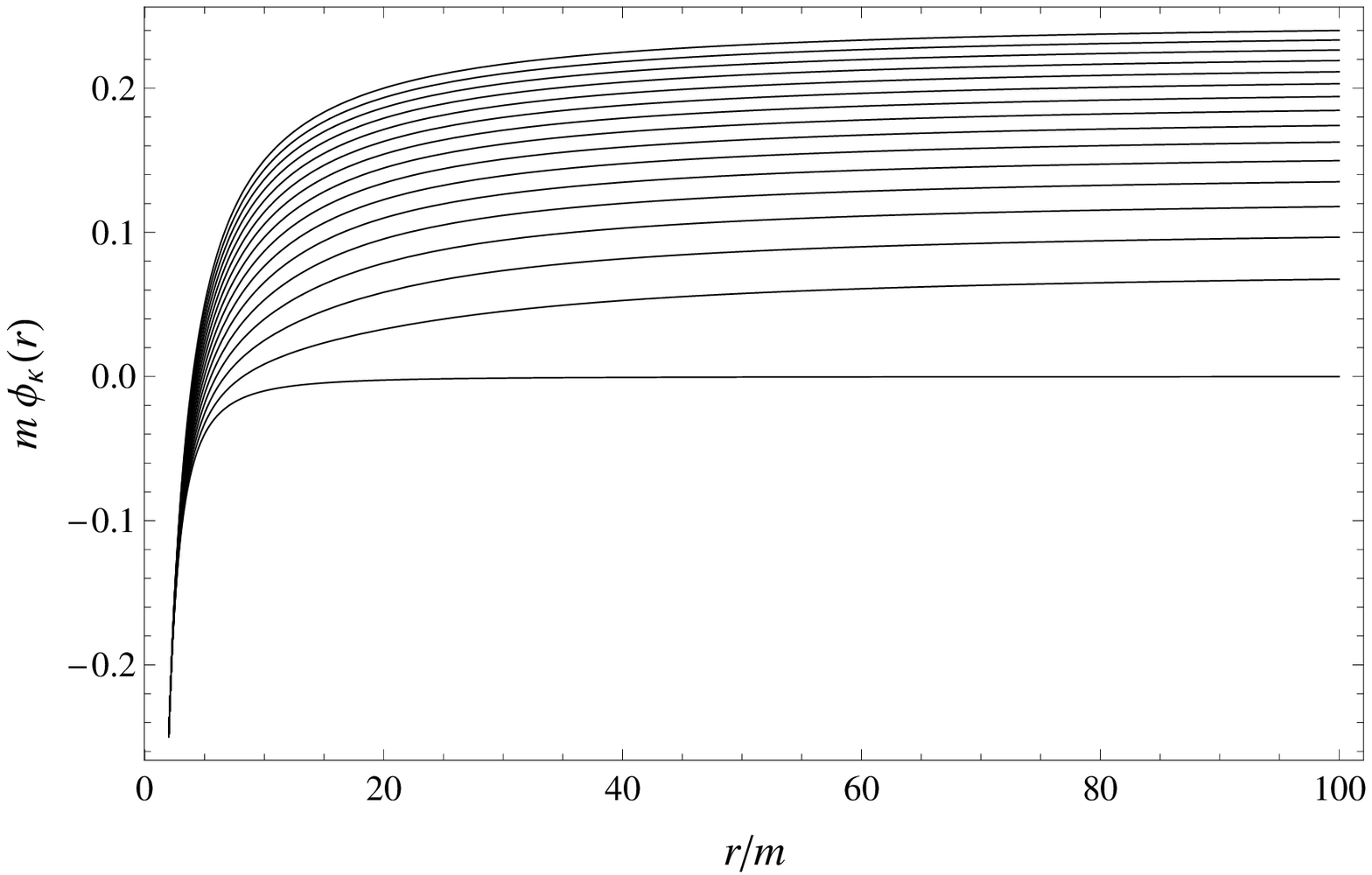}
\includegraphics[width=0.6\linewidth]{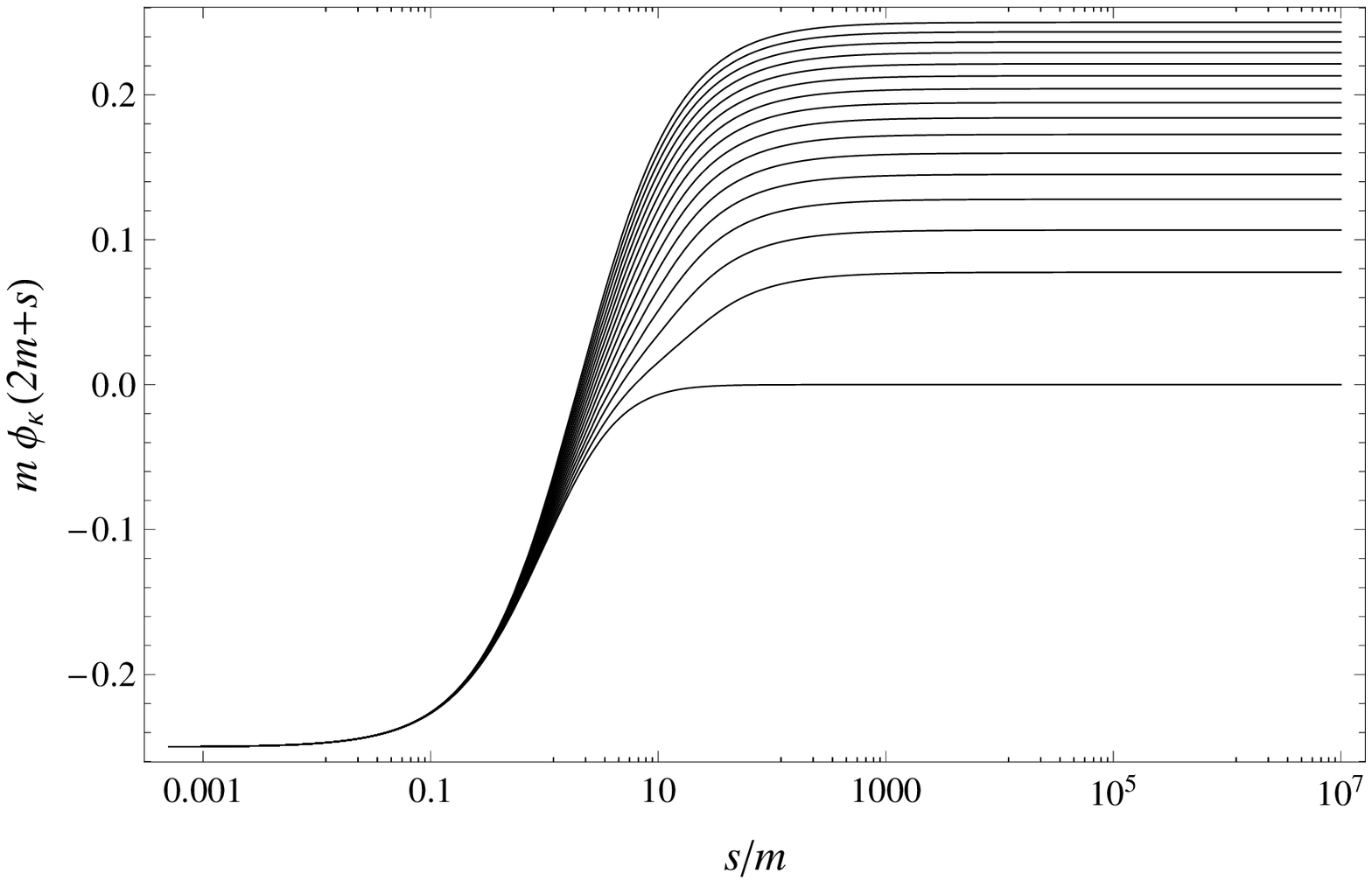}
\caption{Top: Plots of $\phi_\kappa(r)$ as a function of $r$ for evenly spaced values of $\kappa$ going from
$\kappa=-3$ (lowest curve) to $\kappa=-2$ (uppermost curve).
Bottom: Semilog plots of $\phi_\kappa(s+2m)$ as a function of $s$ associated with the same values of $\kappa$ as before.}
\label{fig:phiks}
\end{center}
\end{figure}

Figure~\ref{fig:phiks} shows a representative subset of the family
of solutions $\phi_\kappa$, with $\kappa\in[-3,-2]$ (these plots were generated
as those in Fig.~\ref{fig:superfamilia}).
Notice that it numerically illustrates all the analytical results obtained above.

It follows from Eq.~(\ref{Ckappa}) that, given $R>2m$,
$\frac{m}{R^2}<\phi_\kappa(\infty)-\phi_\kappa(R)<\frac{1}{R}$. Thus
\[
\phi_\kappa(R)+\frac{m}{R^2}<\phi_\kappa(\infty)<\phi_\kappa(R)+\frac{1}{R}
\]
and this leads to accurate estimates for the action via \eq{energia}:
\[
S_\kappa = 1+ 4m\,\phi(\infty).
\]
In fact, data from Fig.~\ref{fig:phiks} yield estimates of $S_\kappa$ with an error of one part in $10^6$.
We note that the above expression, on account of~\eq{slarge.a},
limits $S_\kappa$ to $1\le S_\kappa\le 2$ if $\kappa\in[-3,-2]$.

\begin{figure}[htbp]
\begin{center}
\includegraphics[width=0.6\linewidth]{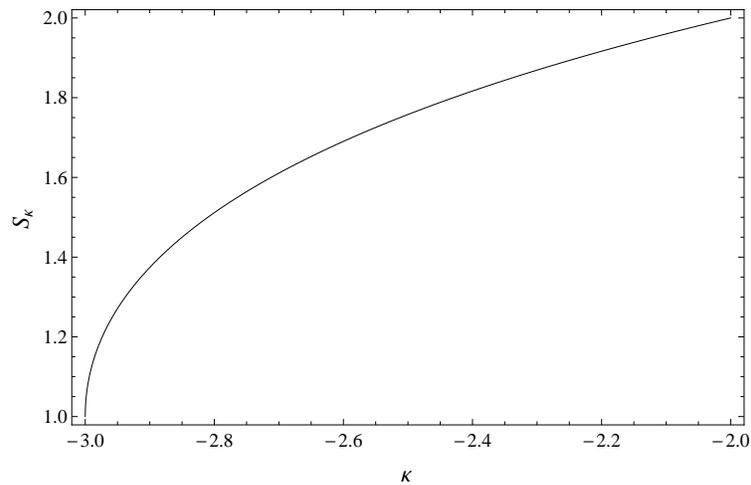}
\caption{Action $S_\kappa$ (associated with the solution $\phi_\kappa$) as a function of $\kappa$.}
\label{fig:energia}
\end{center}
\end{figure}

Figure~\ref{fig:energia} shows how $S_\kappa$ depends on
the parameter $\kappa$ for $-3\le \kappa\le -2$. We see that the action varies continuously  with the parameter $\kappa$.
It is worth noting that the derivative of $S_\kappa$ becomes singular exactly when the physical
character of the solution abruptly changes from a dyon to a monopole, at $\kappa=-3$.


\section{Analytical Expression for the Solution}

So far we have proved the existence of the solutions and presented them numerically.
Now we make use of the mapping method introduced in~\cite{mapping,mapping2}
to obtain an analytical expression for the solution defined for all $r\in[2m,\infty)$.
This method is based on the fact that, if the solution exists and has no singularities in its domain,
then there must be an angular sector of the complex plane containing the domain
in which the function is analytic. If we compactify this region to the complex unit disk,
making a transformation from $z$ to a new variable $\omega$, then there should be a power series expansion for
the solution in $\omega$ which is convergent for $|\omega|<1$. To employ this method, first we use the
shifted variable $s=r-2m$. Then we define a new variable related to the old one
(we are considering the analytic continuation of the solution in the complex plane)
by the following conformal transformation \cite{mapping}
\be
\label{w de z}
s \rightarrow \omega = \frac{(1+s/R)^{1/a}-1}{(1+s/R)^{1/a}+1},
\ee
where $R$ and $a$ are chosen in order to avoid the first singularity of the analytic continuation of $\phi(s)$ in the complex plane.
We found that the choice $R=m$ and $a=1$ apparently gives an infinite radius of convergence for the power series
(evidence for this will be presented in what follows). The inverse transformation is readily obtained
\be
\label{s de w}
w \rightarrow s = R \left[ \left(\frac{1+\omega}{1-\omega}\right)^a -1 \right].
\ee

Using this change of variables in \eq{EDO_phi} yields
\be
\label{EDO_psi_w}
(\omega-1)^2 \omega \frac{d^2\psi}{d\omega^2} + 4 m \psi \frac{d\psi}{d\omega}  - 2\psi =0,
\ee
where $\psi(\omega) = \phi(s(\omega)+2m)$. In this new variable the Charap and Duff solutions become polynomials of degree $1$ and $2$. The family of solutions considered here thus interpolates, in the new variable, between a straight line and a parabola.

We want to write the solutions to \eq{EDO_psi_w} as a power series in $\omega$.
As we have already discussed, we are interested in solutions with $\phi(2m)=\psi(0) = -\frac{1}{4m}$.
Therefore, if we write $\psi(\omega) = \frac{1}{m} \sum_{n=0}^{\infty} b_n \omega^{n+\lambda}$, we have $\lambda = 0$ and $b_0 = - \frac{1}{4}$.
Substituting the series expression in \eq{EDO_psi_w} we see that $b_1 = \frac{1}{2}$, $b_2$ is left free and $b_3=0$.
The relationship between $b_2$ and $\kappa$ is given by $b_2=\frac{\kappa +2}{4}$. All the other coefficients $b_{n+1}$
with $n\geq 3$ can be determined via the relation
\be
\label{recorrencia}
b_{n+1} = \frac{2b_n\left[n(n-1)+1 \right] - b_{n-1}(n-1)(n-2) - 4\sum_{q=0}^{n-1}(q+1)b_{n-q}b_{q+1}}{\left( n^2-1 \right)} .
\ee

After evaluating the parameters $b_n$ we can go back to the original variable $r$:
\be
\label{analytical solution}
\phi_\kappa(r) = \frac{1}{m} \sum_{n=0}^{\infty} b^{(\kappa)}_n \left(1-\frac{2m}{r}\right)^n,
\ee
where $b^{(\kappa)}_n$ depends on $\kappa$ via $b_2$. The first few terms of (\ref{analytical solution}) are displayed below:
\begin{eqnarray*}
\phi_\kappa(r) &=& -\frac{1}{4m} + \frac{1}{2m} \left( 1- \frac{2m}{r} \right) + \frac{\kappa + 2}{4m} \left( 1- \frac{2m}{r} \right)^2 - \frac{\kappa^2 + 5 \kappa + 6}{16 m}\left( 1- \frac{2m}{r} \right)^4 + \\
& & - \frac{\kappa^2 + 5 \kappa + 6}{15 m}\left( 1- \frac{2m}{r} \right)^5 +
\mathcal{O}\left[ \left( 1- \frac{2m}{r} \right)^6 \right].
\end{eqnarray*}
\begin{figure}[htbp]
\begin{center}
\includegraphics[width=0.6\linewidth]{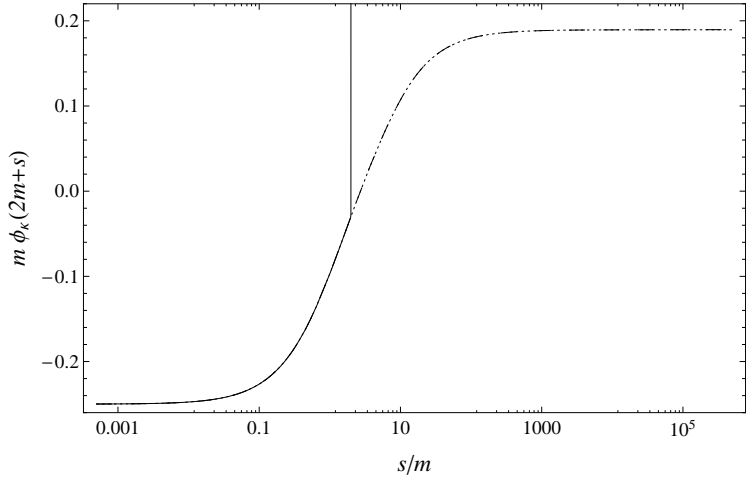}
\caption{Plots of the series expressions obtained before (solid line) and after (dotted line) applying
the mapping method. The dashed line corresponds to the numerical solution obtained in the
previous section. Notice that the series solution given by the mapping method (truncated here at n=30\,000)
is virtually indistinguishable from the numerical solution.
In all the cases, $\kappa=-2.5$.}
\label{fig:serieb}
\end{center}
\end{figure}

Figure~\ref{fig:serieb} shows a typical solution obtained using this procedure and compares it
with its counterparts obtained by the methods of the previous section.
It illustrates the fact that the series solution whose first terms are given by \eq{phi_kappa} has a
radius of convergence not greater than $2m$ and that the series solution obtained
by the mapping method has an apparently infinite radius of convergence (see below). It also shows that the latter is
virtually indistinguishable from the numerical solution obtained in the previous section.

\begin{figure}[htbp]
\begin{center}
\includegraphics[width=0.6\linewidth]{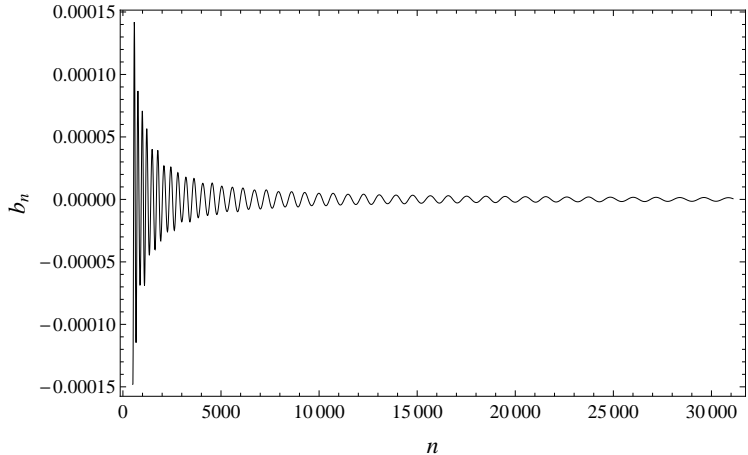}
\caption{Plot of the coefficients $b_n$ as a function of $n$ for $\kappa=-2.5$.}
\label{fig:coef}
\end{center}
\end{figure}

Figure~\ref{fig:coef} shows the behavior of $b_{n}$ for a typical value of $\kappa$. We observe numerically that $|b_n|\rightarrow 0$ as
$n\rightarrow \infty$ so that $|b_n|$ is certainly limited by a constant $c>0$.
On the other hand, the series $\sum_{n=0}^{\infty} c\, \omega^n$ (whose sum  is $c/(1-\omega)$) is absolutely convergent for $|\omega|<1$ and using the comparison test we see that, as far as the above numerical argument is justified, the series solution will be absolutely convergent for $\omega \in [0,1)$, i.e., for any $r$ extending from $2m$ to infinity~\cite{fn5}.

\section{Concluding Remarks}

We have studied, in a systematic way, smooth spherically symmetric self-dual solutions of
Yang-Mills theory in the Euclidean Schwarzschild background.
We showed that our approach recovers all the previously known solutions and leads to a new one-parameter family
of instantons with continuous actions in the range $1< S < 2$.
After studying the global behavior of our solutions, we exhibited them numerically and
employed the mapping method~\cite{mapping,mapping2} to express them analytically.
Finally, we examined the physical properties of this family and showed that it consists of dyons of unit (magnetic and electric) charge
which interpolate between the Charap and Duff's monopole and dyon.
Furthermore, the singularity in the derivative of the action $S_\kappa$ with respect to the parameter $\kappa$ (Fig.~\ref{fig:energia})
signals the abrupt change in the physical character of the solution (from dyons to a monopole).

It is interesting to note that it has been argued, on physical grounds, that no time-dependent solutions
(modulo gauge transformations) exist in Euclidean Schwarzschild space~\cite{tekin}. As a result,
our approach would exhaust all possible smooth spherically symmetric solutions (static or not) in Schwarzschild space.

We end with a more mathematical note. Recent results impose strict constraints on the action spectrum of instantons
defined on ALF geometries~\cite{insgrav}, of which the Euclidean Schwarzschild space is an example.
In particular, it is possible to show that an instanton defined on the Euclidean Schwarzschild geometry
will necessarily have integer action if it satisfies a certain rapidly decaying
condition, which means that the curvature decays faster than $1/r^2$ as $r\to\infty$.
We note that the noninteger solutions presented in this paper do not satisfy this assumption, see Eq~(\ref{asymptotic}), so no contradiction arises here. On the contrary, this discussion shows that the hypotheses of~\cite{insgrav} are, in a sense, as weak as possible.


\acknowledgments

RAM is indebted to G. Etesi for calling his attention to the subject of instantons
in ALF spaces, particularly in the Euclidean Schwarzschild geometry, and for
many helpful discussions. The authors thank M. Jardim for useful discussions.
This work was supported by FAPESP and CNPq.


\appendix

\section{Global analysis of the solutions}

We show here that~\eqs{slarge} hold whenever $\kappa_1<\kappa_2\le-2$.
We start by noting that, if $\kappa<-2$, then
\be
\label{lema:2}
\phi'_\kappa(r) < \frac{1}{r^2} = \phi'_{-2}(r)  \quad \forall r\in (2m,\infty).
\ee
This condition is clearly true for $r-2m \ll 1$ (see~\eq{ssmall.a}). Moreover, if this condition ceased to hold there would be a $r_0$ such that $\phi'_\kappa(r_0) = \tfrac{1}{r_0^2}$. But then, by the uniqueness theorem for ODEs, we would have $\phi_\kappa(r)= c-1/r$ for some $c$ which, together with the condition on $r=2m$, leads to $\phi_\kappa(r)=\phi_{-2}(r)$ and therefore $\kappa=-2$, contradicting our assumption.

Now we prove that, if $\kappa_1<\kappa_2\le-2$, then
\be
\label{prop}
\phi'_{\kappa_1}(r) < \phi'_{\kappa_2}(r) \quad \forall r\in (2m,\infty).
\ee
As already discussed, this holds for $r-2m \ll 1$ (see \eq{ssmall.b}). Suppose that there exists $r\in(2m,\infty)$ such that
$\phi'_{\kappa_1}(r) = \phi'_{\kappa_2}(r)$ and let $r_0$ be the least value
of $r$ with this property. Since $\phi'_\kappa$ is continuous
(due to the general theory of ODEs),
we have
\be
\phi'_{\kappa_1}(r) < \phi'_{\kappa_2}(r) \quad \forall r\in (2m,r_0).
\label{aux}
\ee
Integrating both sides of~(\ref{aux}) from $2m$ to $r\in  (2m,r_0]$ then leads to
\be
\phi_{\kappa_1}(r)  < \phi_{\kappa_2}(r) \quad \forall r\in (2m,r_0],
\label{eqaux}
\ee
since $\phi_\kappa(2m)=-1/4m$ for all $\kappa$.
On the other hand, \eq{EDO_phi} yields
\be
\label{ed_dif}
\frac{1}{2}(r_0-2m)(\phi''_{\kappa_2}-\phi''_{\kappa_1})(r_0)+
r_0 \left( \phi'_{\kappa_1}(r_0)-\frac{1}{r_0^2} \right)
(\phi_{\kappa_2}-\phi_{\kappa_1})(r_0) =0,
\ee
since we assumed that $(\phi'_{\kappa_1}-\phi'_{\kappa_2})(r_0)=0$.
It follows from Eqs.~(\ref{lema:2}) and~(\ref{eqaux}) that
$(\phi''_{\kappa_2}-\phi''_{\kappa_1})(r_0)>0$, i.e., $\frac{d}{dr}(\phi'_{\kappa_2}-\phi'_{\kappa_1})(r_0)>0$.
In this way, $\phi'_{\kappa_2}(r_0-\epsilon)<\phi'_{\kappa_1}(r_0-\epsilon)$
for $\epsilon$ positive and sufficiently small. But this contradicts~\eq{eqaux}.
Therefore, no such $r_0$ exists and \eq{prop} follows.

Finally, integrating \eq{prop} from $2m$ to $r$ we see that, for $\kappa_1<\kappa_2\le-2$,
\be
\label{cor:mon}
\phi_{\kappa_1}(r) < \phi_{\kappa_2}(r)  \quad \forall r\in (2m,\infty).
\ee
This concludes the proof.



\begin{thebibliography}{99}
\bibitem{fn1}
As will be discussed later
the solutions we find are, in fact, time-dependent when expressed in the gauge where they are regular,
but this presents no problem to the procedure just described.
\bibitem{fn2}
A purely imaginary $\alpha$ would give rise to a $\so(2,1)$-valued gauge potential in (\ref{ansatz}) (see also~\cite{mj}).
\bibitem{fn3}
We note that it is not possible to start the numerical integration
from $r=2m$, since this is a singular point of the differential equation
under consideration.
\bibitem{fn4}
Clearly, since (\ref{EDO_phi}) is a second order ODE, this property does not hold for
generic solutions which do not belong to this family.
\bibitem{fn5}
It is is worth noting that the numerical approach may still be of practical relevance
since the convergence rate of the series solution becomes quite slow for $\omega$ very close to~$1$.
\end{thebibliography}
\end{document}